\documentclass[final,5p,times,twocolumn,sort&compress]{elsarticle}
\usepackage{graphicx}
\usepackage{amsmath}
\usepackage{multirow}
\usepackage{amssymb}

\begin{document}

\begin{frontmatter}

\title{Final state effects on charge asymmetry of pion elliptic flow \\in high-energy heavy-ion collisions}

\author{Guo-Liang Ma}  
            
\address{Shanghai Institute of Applied Physics, Chinese
Academy of Sciences, Shanghai 201800, China}


\begin{abstract}

Within a multi-phase transport (AMPT) model with the string melting mechanism and imported initial electric quadrupole moment, the difference between the elliptic flow of positive and negative pions is calculated. The slope parameter $r$ of the linear dependence of $\Delta v_{2}=v_{2}(\pi^{-})-v_{2}(\pi^{+})$ on $A_{ch}=(N^{+}-N^{-})/(N^{+}+N^{-})$ is yielded owing to the parton cascade, which converts the initial electric quadrupole distribution into the final charge-dependent elliptic flow. The slope parameter $r$ is found to be increased by the hadronization given by the coalescence, and decreased by the resonance decays. The slope parameter $r$ is very sensitive to both the initial electric quadrupole percentage and centrality bin and consequently a helpful constraint on the quadrupole moment of the chiral magnetic wave is obtained for Au+Au collisions at the top RHIC energy.

\end{abstract}


\end{frontmatter}

\vspace{-1pc}
\section{Introduction}
\label{sec:intro}

High-energy heavy-ion collisions provide a good chance to study many novel phenomena for the strongly interacting matter under a hadron-scale magnetic field~\cite{Kharzeev:2013ffa,Liao:2014ava}. Recently, it has been proposed that a gapless chiral magnetic wave (CMW) could be formed by the interplay between the chiral magnetic effect (CME)~\cite{Kharzeev:2004ey,Kharzeev:2007tn,Kharzeev:2007jp} and the chiral separation effect (CSE)~\cite{Son:2004tq,Metlitski:2005pr} along the direction of the magnetic field in the quark-gluon plasma (QGP) owing to the triangle anomaly of QCD~\cite{Kharzeev:2010gd}. The CMW leads to an electric charge quadrupole moment in the initial coordinate space of the QGP which can be finally translated into a charge asymmetry of the elliptic flow in the final momentum space of pions via the collective expansion~\cite{Burnier:2011bf, Burnier:2012ae}. A possible signal of the CMW has been observed through a linear charge asymmetry [$A_{ch}=(N^{+}-N^{-})/(N^{+}+N^{-})$] dependence of the elliptic flow difference between positive and negative pions [$\Delta v_{2}=v_{2}(\pi^{-})-v_{2}(\pi^{+})$], i.e. $\Delta v_{2}=rA_{ch}+\Delta v_{2}(0)$, in the recent STAR measurement~\cite{Wang:2012qs,Ke:2012qb}. Hydrodynamics with triangle anomalies~\cite{Son:2009tf} has been applied in numerical studies of the CMW by several theory groups. Hongo $et~al.$ quantitatively verified that the intercept $\Delta v_2(0)$, rather than the slope parameter $r$, is a sensitive signal of the anomalous transport~\cite{Hongo:2013cqa}. However, Yee and Yin argued that the treatment of freeze-out condition is crucial to identify the CMW contribution to the slope parameter $r$ due to a freeze-out hole effect~\cite{Yee:2013cya}. Taghavi and Wiedemann thought that the asymmetry in the electric charge distribution induced by the CMW is too small to be experimentally accessible unless a sizable axial charge asymmetry already exists in the initial condition~\cite{Taghavi:2013ena}. Therefore, the key issue in the debate is whether the chiral anomaly can lead to an experimentally accessible signal through the final state interactions in high-energy heavy-ion collisions. On the other hand, the chiral kinetic theory has been developed based on magnetic monopoles in the momentum space with the corresponding Berry curvature flux~\cite{Chen:2012ca,Son:2012zy}. However, to my knowledge, so far the CMW theory has not been incorporated with a dynamical transport model with the final state interactions. In this work, the AMPT model with an imported initial electric quadrupole moment is implemented to study the final state effects on the charge asymmetry of the pion elliptic flow in high-energy heavy-ion collisions. It is found that the final state interactions play an important role in the measured charge asymmetry of the pion elliptic flow, which provides a helpful constraint on the quadrupole effect of the CMW.

The paper is organized as follows. In Section \ref{sec:model} I give a brief introduction to the AMPT model with an initial electric quadrupole distribution. Results and discussion about the final state effects on the charge asymmetry of the pion elliptic flow are presented in Section \ref{sec:resul1}. Finally, a summary is given in Section \ref{sec:concl}.

\vspace{-1pc}
\section{Model introduction}
\label{sec:model}

The AMPT model with the string melting mechanism~\cite{Lin:2004en} is employed to study the effects of the final state interactions on charge asymmetry of the pion elliptic flow. It consists of four main stages of high-energy heavy-ion collisions: The initial condition, parton cascade, hadronization, and hadronic rescatterings. The initial condition, which includes the spatial and momentum distributions of minijet partons and soft string excitations, is obtained from the HIJING model~\cite{Wang:1991hta,Gyulassy:1994ew}. The parton cascade starts the parton evolution with a quark-anti-quark plasma from the melting of strings, during which only elastic partonic interactions are included at present~\cite{Zhang:1997ej}. After the partons freeze-out dynamically, they are recombined into hadrons via a simple coalescence model~\cite{Lin:2001zk}. Dynamics of the subsequent hadronic matter is then described by a relativistic transport (ART) model~\cite{Li:1995pra}. Because the current implementation of the ART model does not conserve the electric charge, only resonance decays are considered and hadronic scatterings are turned off to ensure charge conservation in this study. The AMPT model with the string melting mechanism has been shown to reproduce well the elliptic flow data at RHIC~\cite{Chen:2004dv, Zhang:2005ni, Chen:2006ub,Han:2011iy} and the LHC~\cite{Xu:2011fi,Xu:2011jm}. Recently, the AMPT model has been implemented to study charge azimuthal correlations $\left\langle\cos(\phi_{\alpha} \pm \phi_{\beta})\right\rangle$ and it was concluded that the final state interactions can play a significant role in affecting the initial electric dipole moment from the CME in Au+Au collisions at the top RHIC energy~\cite{Ma:2011uma}.

In order to effectively simulate the CMW effect with an electric quadrupole distribution, the position coordinates ($x, y, z$) of a fraction of the small-$|y|$ $u$ (or $\bar{d}$) quarks are switched with those of large-$|y|$ $\bar{u}$ (or $d$) quarks in the initial states of the AMPT events with $A_{ch}>$-0.01, while a contrary manner is applied for the AMPT events with $A_{ch}<$-0.01. Only two quark flavors ($N_{f}$=2) are considered for simplicity. The coordinate system is set up so that the $x$-axis is in the reaction plane and the $y$-axis is perpendicular to the reaction plane with the $z$-axis being the incoming direction of one nucleus. The fraction is represented by a relative percentage with respect to the total number of quarks. The $A_{ch}$ intercept cut-off -0.01 is chosen based on the fact that $v_{2}$ difference between $\pi^{-}$ and $\pi^{+}$ changes its sign around $A_{ch}\sim$ -0.01, as seen in the STAR preliminary data~\cite{Wang:2012qs,Ke:2012qb}. It should be emphasized that since there is no electromagnetic field and the chiral charge degrees of freedom in the AMPT model, the transport development of the CMW-induced electric quadrupole distribution after a rapid decay of electromagnetic field is the main interest of this work. Consistently with the previous AMPT studies at the RHIC energies, a large partonic interaction cross section, 10 mb, is chosen to simulate Au+Au collisions at $\sqrt{s_{_{\rm NN}}}$ = 200 GeV.

\vspace{-1pc}
\section{Results and discussion}
\label{sec:resul1}

\begin{figure}
\includegraphics[scale=0.45]{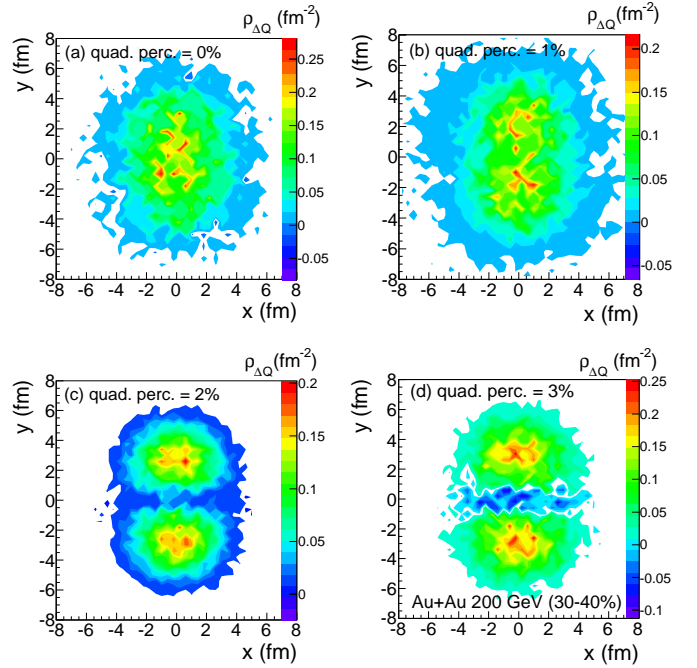}
\caption{(Color online) The mid-pseudorapidity net electric charge density in the transverse plane in the initial partonic state with different initial quadrupole percentages [(a)-(d)] for the centrality bin of 30-40\% in Au+Au collisions at $\sqrt{s_{_{\rm NN}}}$ = 200 GeV.}
 \label{fig-xydis}
\end{figure}

Figure~\ref{fig-xydis} (a)-(d) shows the contour plots of mid-pseudorapidity ($|\eta|<$ 1) net electric charge density $\rho_{\Delta Q} = d\Delta Q/dxdy$ in the transverse plane ($z$-axis-integrated) in the initial state with different initial quadrupole percentages for the centrality bin of 30-40\% in Au+Au collisions at $\sqrt{s_{_{\rm NN}}}$ = 200 GeV. An electric quadrupole distribution, which is characterized as an additional negatively charged quarks around the equator and additional positively charged quarks near two poles, becomes more and more distinct with the increasing of the initial quadrupole percentage.

\begin{figure}
\includegraphics[scale=0.45]{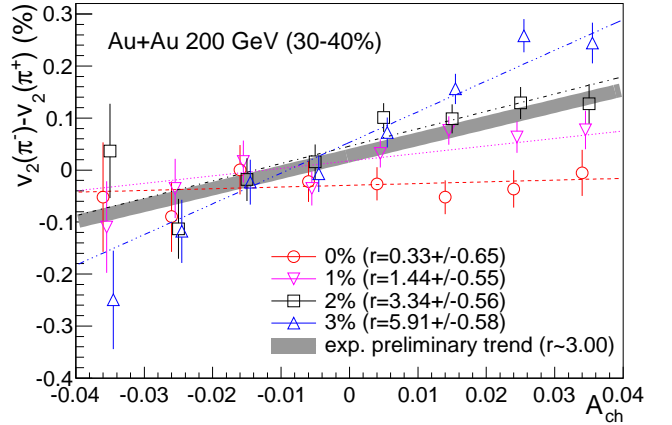}
\caption{(Color online) $v_{2}$ difference between $\pi^{-}$ and $\pi^{+}$ as a function of charge asymmetry $A_{ch}$ with different initial quadrupole percentages (open symbols) for the centrality bin of 30-40\% in Au+Au collisions at $\sqrt{s_{_{\rm NN}}}$ = 200 GeV, where the different types of lines are the corresponding linear fits and the band represents the experimental preliminary trend~\cite{Wang:2012qs,Ke:2012qb}. Some points are slightly shifted for clarity.
}
 \label{fig-Deltav2vsAchfordiffquadPercents}
\end{figure}

Since the final state interactions can convert the initial geometry eccentricity into the final elliptic flow in non-central heavy-ion collisions, an initial electric quadrupole distribution is believed to result in a charge asymmetry of the final pion elliptic flow according to the CMW theory.  Figure~\ref{fig-Deltav2vsAchfordiffquadPercents} shows the elliptic flow difference between $\pi^{-}$ and $\pi^{+}$ (0.15 $< p_{T}<$ 0.5 GeV/$c$) as functions of the charge asymmetry factor $A_{ch}$ for different initial quadrupole percentages for the centrality bin of 30-40\% in Au+Au collisions at $\sqrt{s_{_{\rm NN}}}$ = 200 GeV, where $A_{ch}=(N^{+}-N^{-})/(N^{+}+N^{-})$, and $N^{\pm}$ is the number of positively or negatively charged particles ($p_{T} >$ 0.15 GeV/$c$ and $|\eta| <$ 1). The slope parameter $r$ can be extracted through a linear fitting of $\Delta v_{2}=rA_{ch}+\Delta v_{2}(0)$, where $r$ is thought to be related to $2q_{e}/\rho_{e}$, the twofold ratio of quadrupole moment to net charge density from the CMW theory. The slope parameter $r$ is consistent with zero for the AMPT result with zero initial quadrupole percentage, which is consistent with an UrQMD calculation~\cite{Wang:2012qs,Ke:2012qb} but different from a hydrodynamic calculation with late local charge conservation at freeze-out~\cite{Bzdak:2013yla}. The slope parameter $r$ increases with the increasing initial quadrupole percentage. The AMPT result with an initial quadrupole percentage of 2\% can basically describe the experimental preliminary trend with the slope parameter $r\sim3$. These results are different from the recent numerical simulations from anomalous hydrodynamics, in which the slope parameter $r$ is not sensitive to anomalous transport effects~\cite{Hongo:2013cqa}. The AMPT results indicate that the charge asymmetry of the pion elliptic flow can be induced by an initial electric quadrupole distribution.

\begin{figure}
\includegraphics[scale=0.45]{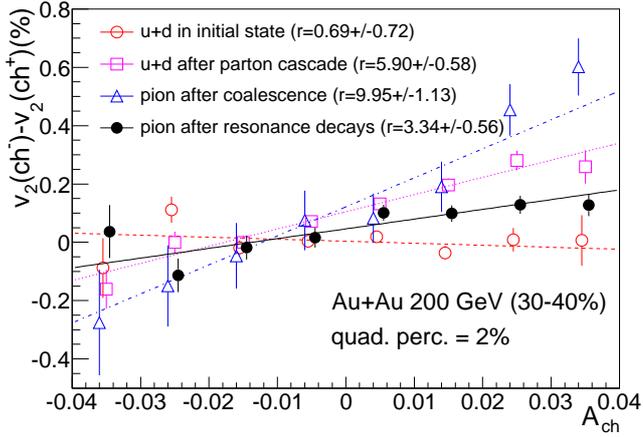}
\caption{(Color online) $v_{2}$ difference between negatively and positively charged particles as a function of charge asymmetry $A_{ch}$ for different stages of a heavy-ion collision, where the different types of lines are the corresponding linear fits. Here an initial quadrupole percentage is 2\% for the centrality bin of 30-40\% in Au+Au collisions at $\sqrt{s_{_{\rm NN}}}$ = 200 GeV. Some points are slightly shifted for clarity.
}
 \label{fig-diffstages}
\end{figure}

Because a heavy-ion collision is actually a dynamical process which involves many important evolution stages, it is essential to see how various stages affect the charge asymmetry of the pion elliptic flow. Figure~\ref{fig-diffstages} shows the charge asymmetry $A_{ch}$ dependence of $v_{2}$ differences between negatively and positively charged particles for different evolution stages with an initial quadrupole percentage of 2\% and for the centrality bin of 30-40\% in Au+Au collisions at $\sqrt{s_{_{\rm NN}}}$ = 200 GeV. In the initial state, the slope parameter $r$ is close to zero, since $v_{2}$ for both negatively charged quarks [$v_{2}(\bar{u}+d)$] and positively charged quarks [$v_{2}(u+\bar{d})$] is consistent with zero as no final state interactions happen at this point. However, an $A_{ch}$ linear dependence of $\Delta v_{2}$ for quarks is formed with the slope parameter $r$ of 5.90 $\pm$ 0.58 after the process of the parton cascade with frequent partonic interactions, which indicates that the initial electric quadrupole distribution has been transferred into the charge asymmetry of the quark elliptic flow. The slope parameter $r$ for pions is increased to 9.95 $\pm$ 1.13 after the hadronization in which $u$ ($\bar{u}$) and $\bar{d}$ ($d$) quarks are recombined into positively (negatively) charged pions. It is consistent with the hadronization of Cooper-Frye freeze-out in anomalous hydrodynamical simulations, which is thought as a dominant source of the CMW contribution to the slope parameter $r$ due to a freezeout hole effect~\cite{Yee:2013cya}. However, the slope parameter $r$ for pions is reduced to 3.34 $\pm$ 0.56 after the process of resonance decays since many secondary pions born from resonance decays, e.g. $\rho$ decays, can smear or destroy the charge asymmetry of the primary pion elliptic flow. However, since the smearing effect of the resonance decays should be more effective for low-$p_{T}$ range, one should see a clearer charge asymmetry of the pion elliptic flow by increasing the $p_{T}$ cut of pions. Therefore, the measured slope parameter $r$ is actually an end product from the dynamical evolution in high-energy heavy-ion collisions.

\begin{figure}
\includegraphics[scale=0.45]{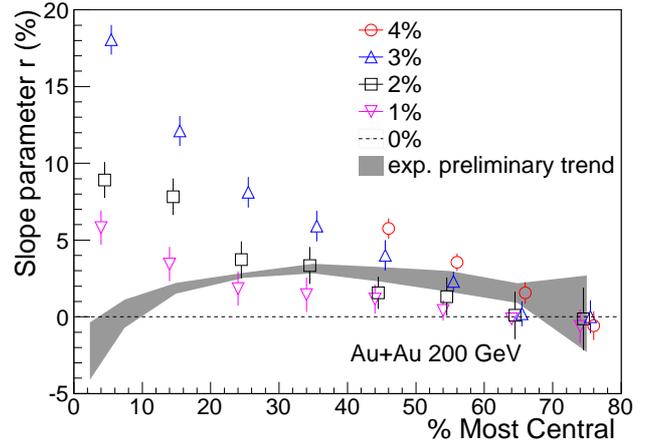}
\caption{(Color online)  The slope parameter $r$ as a function of the centrality bin in Au+Au collisions at $\sqrt{s_{_{\rm NN}}}$ = 200 GeV. The open symbols represent different initial quadrupole percentages in the AMPT simulations. The dash line represents r = 0, corresponding to vanishing quadrupole percentage. The band represents the experimental preliminary trend~\cite{Wang:2012qs,Ke:2012qb}. Some points are slightly shifted for clarity. }
 \label{fig-slopes}
\end{figure}

\begin{table*}
\caption{The estimations of initial electric quadrupole percentages for different centrality bins in Au+Au collisions at $\sqrt{s_{_{\rm NN}}}$ = 200 GeV.}
\centering
\begin{tabular}{ccccccccc}
\hline 
Centrality bin & 0-10\% & 10-20\% & 20-30\% & 30-40\% & 40-50\% & 50-60\% & 60-70\% & 70-80\%  \\
\hline  
Initial quadrupole percentage &0\% & $<$1\% &1-2\% &1-2\% &2-3\% &2-3\% &3-4\% &? \\
\hline 
\end{tabular}
\label{tab:example}
\end{table*}

Figure~\ref{fig-slopes} shows the centrality dependence of the slope parameter $r$ for different initial quadrupole percentages in Au+Au collisions at $\sqrt{s_{_{\rm NN}}}$ = 200 GeV, where only four peripheral centrality bins are shown for the initial quadrupole percentage of 4\%. The slope parameter $r$ is sensitive to both initial quadrupole percentage and centrality bin, i.e. it increases with the initial quadrupole percentage and decreases from central to peripheral collisions. By comparing the AMPT results with the experimental preliminary trend, a helpful constraint on the centrality dependence of the initial quadrupole percentage can be estimated, as listed in Table~\ref{tab:example}. For the most central bin of 0-10\%, the experimental trend favors no initial quadrupole percentage, which is consistent with the CMW expectation that the strength of electromagnetic field is relatively small due to a small number of the spectator nucleons in the most central collisions. The initial quadrupole percentage increases from central to peripheral collisions except the most peripheral bin of 70-80\%, which supports that a more quadrupole deformation is created with the increasing strength of the electromagnetic field from central to peripheral collisions. Unfortunately, the increase of initial quadrupole percentage cannot result in an increase of the slope parameter $r$ owing to the decrease of the strength of partonic interactions from central to peripheral collisions. At the most peripheral bin of 70-80\%, the initial quadrupole percentage is uncertain from the AMPT simulations because the partonic evolution is too short or weak to convert any finite initial electric quadrupole moment into a charge asymmetry of the pion elliptic flow. 

The extracted initial electric quadrupole percentages (or distributions) are consistent with the numerical solutions of the CMW equation which generates electric quadrupole deformations~\cite{Kharzeev:2010gd,Burnier:2011bf}. However, it is worth noticing that the initial electric quadrupole distribution is introduced around a mean formation time of partons \footnote{I checked that it is around 2 fm/$c$ for Au+Au collisions at $\sqrt{s_{_{\rm NN}}}$ = 200 GeV.} in this study, when the magnetic field is expected to vanish away. Since the lifetime of a magnetic field in the plasma remains unclear so far, it would be interesting to study the time dependence of the initial electric quadrupole distribution to further constrain the lifetime of magnetic field in future investigations.

\vspace{-1pc}
\section{Summary}
\label{sec:concl}

In summary, an $A_{ch}$-dependent charge asymmetry of the pion elliptic flow is reproduced by introducing an initial electric quadrupole moment into the initial state of the AMPT model with the string melting mechanism. The process of parton cascade is essential for the formation of the slope parameter $r$ of the linear dependence of $\Delta v_{2}$ on $A_{ch}$, but other evolution stages such as coalescence and resonance decays also affect the slope parameter $r$. Because the slope parameter $r$ is very sensitive to the initial electric quadrupole percentage and centrality bin, it provides a helpful constraint on the quadrupole effect from the chiral magnetic wave.

\vspace{-1pc}
\section*{Acknowledgements}
I thank Adam Bzdak, Dmitri Kharzeev, Jinfeng Liao, Aihong Tang, Bin Zhang for helpful discussions. This work was supported by the Major State Basic Research Development Program in China under Grant No. 2014CB845404, the NSFC of China under Grants No. 11175232, No. 11035009, and No. 11375251, the Knowledge Innovation Program of CAS under Grant No. KJCX2-EW-N01, the Youth Innovation Promotion Association of CAS under Grant No. Y329051012, the project sponsored by SRF for ROCS, SEM, CCNU-QLPL Innovation Fund under Grant No. QLPL2011P01, and the ``Shanghai Pujiang Program" under Grant No. 13PJ1410600.


\end{document}